# Fundamentals of Atmospheric Physics for Engineering

*Cionco, Rodolfo G.*
*Universidad Tecnológica Nacional, San Nicolás, 2900, Argentina*
*gcionco@frsn.utn.edu.ar*

**Abstract**
We present the proposal of an elective for engineering courses, designed to train professionals with a solid foundation in Physics of the Atmosphere interested in environmental and sustainability issues broadly. We propose four chapters that contain a variety of topics but strongly interrelated, which correspond to three main areas: nature of the atmosphere and meteorology relevant to contaminant transport, the dispersion of air pollutants and climate in general. We conclude that it is possible train engineers who understand the basic mechanisms that led to the current atmosphere, atmospheric processes related to local and global climate, the dispersion of air pollutants and key concepts such as solar activity, climatic change and climatic variability, even in one semester. It also discusses the relationship with other subjects and proposes and illustrates a method of course approval based on the performance of work directly applicable to engineering problems.

# Fundamentos de Física de la Atmósfera para carreras de Ingeniería

**Resumen**
Se presenta la propuesta de una materia optativa para carreras de ingeniería, destinada a formar profesionales con sólidos fundamentos en Física de la Atmósfera, interesados en problemáticas medioambientales y de sostenibilidad en sentido amplio. Se proponen cuatro capítulos que contienen temas variados pero fuertemente relacionados entre sí, que se corresponden con tres grandes áreas temáticas referidas a la atmósfera terrestre: naturaleza y meteorología relevante al transporte de contaminantes; dispersión de contaminantes atmosféricos y clima en general. Se concluye que es posible, aun en un cuatrimestre, formar ingenieros que comprendan los mecanismos básicos que dieron origen a la atmósfera actual, los procesos atmosféricos locales y globales relacionados con el clima, la dispersión de contaminantes y conceptos clave como actividad solar, cambio y variabilidad climática. Se analiza brevemente la articulación con otras materias y se propone un método de aprobación de cursada basado en la realización de trabajos de aplicación directa a problemas de ingeniería.
**Palabras claves:** Medioambiente, sostenibilidad, atmósfera, Ingeniería Ambiental.

## 1. INTRODUCCIÓN

Controlar y mitigar el impacto de los ingenios humanos en los ecosistemas naturales y la sociedad toda, es una tarea que exige cada vez más responsabilidad y nivel de conocimientos en todas las ramas de las ingenierías. Satisfacer las necesidades humanas sin causar daño crónico al medioambiente, es decir, *sin sacrificar las posibilidades de las generaciones futuras para atender sus propias necesidades*, es el gran reto para este siglo (1). Generar un mundo *sostenible* es una empresa formidable que requiere un aumento sustancial en el entendimiento del sistema integrado *sociedad-naturaleza* y las alteraciones (comúnmente llamadas *"antropogénicas"*[1]) que los seres humanos les producen al Planeta. El desafío de educar a las generaciones futuras en temas de impacto ambiental y la formación en sostenibilidad, implica un compromiso multidisciplinar que incluye la comprensión del medioambiente, los ecosistemas naturales y su interconexión con el mundo social y económico, y la viabilidad a largo plazo entre ellos.

La sostenibilidad es mucho más que una meta a alcanzar, se ha ido erigiendo como una verdadera disciplina con objetos de estudio y de investigación. Los temas y conceptos fundamentales que subyacen en la teoría y práctica de la sostenibilidad, como materia científico-tecnológica, incluyen la migración de fuentes de energía no-renovables a renovables, la dinámica de sistemas complejos, el comportamiento emergente, los procesos multi-escala, tanto como la vulnerabilidad y capacidad adaptiva entre sistemas. Una meta importante en la investigación en sostenibilidad, es entender cómo patrones y procesos a escala local y regional están influenciados, y a su vez se retroalimentan, de procesos que se manifiestan a escala global en períodos más largos de tiempo (2). Adentrarse en estos problemas, es enfrentar diversas disciplinas en un complejo panorama científico-tecnológico.

Un primer paso, lo constituye el estudio de la atmósfera terrestre. Junto con los océanos, la atmósfera es la vía más universal de interconexión planetaria y uno de los medios más eficaces de transporte de contaminantes. Ciertamente es la vía más eficiente en el transporte de contaminantes entre complejos industriales suburbanos y los núcleos poblacionales más importantes. La atmósfera posee además la propiedad de producir "teleconexión", i.e., conexión a distancia entre procesos que aparentemente no manifiestan una vinculación directa o evidente entre sí. La atmósfera pone en contacto a todo el Globo y a todos los procesos de superficie que en él ocurren. Por lo tanto, la propuesta de una materia dedicada a la física atmosférica y a su injerencia en las problemáticas ambientales es una contribución fundamental para alumnos de ingeniería que deseen profundizar en estos temas particulares o adentrarse en cuestiones más relacionadas con la Ingeniería Ambiental y la sostenibilidad en sentido amplio (i.e., sostenibilidad como disciplina inherente a las prácticas tecnológicas y su relación con el mundo social y natural, más que referida a un proceso técnico puntual, (1)).

El objetivo de este trabajo es dar a conocer la propuesta de una materia de tipo optativo para carreras de ingeniería, destinada a formar profesionales con sólidos fundamentos en Física de la Atmósfera, interesados en problemáticas medioambientales y de sostenibilidad en sentido amplio. El curso se implementará a partir del año 2013 en la Regional San Nicolás de la Universidad Tecnológica Nacional y se origina en el proyecto "Forzantes Externos al Planeta y Variabilidad Climática" homologado por dicha Institución. A pesar de que en algunos países las Ciencias de la Atmósfera se incluyen dentro de la Ingeniería (como en Uruguay), o se exigen en forma obligatoria materias dedicadas a la protección del medioambiente, no es de nuestro conocimiento la existencia de una materia con estas características en el campo de las ingenierías dentro de la República Argentina. Proponemos 4 capítulos que contienen temas variados pero fuertemente relacionados entre sí, que se corresponden con tres grandes áreas temáticas: naturaleza de la atmósfera terrestre y meteorología relevante al transporte de contaminantes; dispersión de contaminantes atmosféricos, y clima en general. Concluimos que es posible, aun en un cuatrimestre, formar ingenieros que comprendan los procesos básicos que dieron origen a la atmósfera actual (muchos de ellos catastróficos), los procesos atmosféricos locales y globales relacionados con el clima, la dispersión de contaminantes en el aire, y conceptos clave como actividad solar, cambio y variabilidad climática. Analizamos brevemente la articulación con otras materias y proponemos y ejemplificamos un método de aprobación de cursada basado en la realización de trabajos de aplicación directa a problemas de ingeniería.

---

[1] Antropogénico parece no ser una buena castellanización del inglés *"antropogenic"*, ya que *génico* se refiere a los genes y *antropo* al hombre, o sea, "genes humanos" y no "de origen humano".

## 2. Marco Teórico

A diferencia de los planetas gigantes (Júpiter, Saturno Urano y Neptuno), las atmósferas de los planetas terrestres parecen haberse formado en un proceso dual, por desgasificación del manto sólido en formación, y por un bombardeo más o menos discontinuo de asteroides o embriones planetarios con alto contenido de elementos volátiles (3). Luego, el material primigenio se fue procesando entre cambios químicos e impactos catastróficos tardíos involucrando grandes cuerpos proto-planetarios (ver (4) para una referencia en castellano). Posteriormente, el desarrollo de la vida a fines del Precámbrico (más o menos antes de que se diversificaran las formas primigenias de vida, hace unos 2 Giga-años), produjo una atmósfera rica en oxígeno y pobre en dióxido de carbono ($CO_2$, actualmente el porcentaje en volumen de $CO_2$ es de 0,03), con cantidades inusualmente altas de metano, a diferencia de lo observado en los otros Terrestres. El océano y los movimientos orográficos tuvieron gran importancia en el reciclado del carbono: grandes cantidades de $CO_2$ emitidas por volcanes fueron "lavadas" de la atmósfera por las lluvias, arrastradas por la corteza terrestre para formar luego rocas calcáreas, las cuales, debido a movimientos tectónicos, terminaban formando parte del magma, posteriormente fundidas y "desgasificadas", emitiéndose nuevamente a la atmósfera por las chimeneas volcánicas, para reiniciar de esta forma, el *ciclo del $CO_2$*.

Actualmente existe una tendencia importante que adjudica al $CO_2$ y otros gases llamados "de invernadero" una importancia trascendental para explicar el calentamiento global observado y medido desde el siglo XIX, debido a su aumento por causas artificiales (industrialización, etc.), y está basada en las publicaciones del Intergovernmental Panel of Climate Change (5). La física detrás de esto sería el comúnmente llamado "efecto invernadero". El "efecto invernadero" es una suposición fenomenológica, que está basada en trabajos antiguos de Arrhenius, Tyndall y Fourier (6, 7); se trata más bien de una conjetura, sin demostración, al menos basada en primeros principios (6). Es más, inicialmente la teoría del $CO_2$ fue propuesta para explicar "enfriamientos" (7). Más allá de cómo se resuelvan éstas y otras controversias detrás del calentamiento global observado o C*ambio Climático* (8), lo cierto es que las cantidades de $CO_2$ atmosférico tampoco han permanecido constantes a lo largo de los tiempos geológicos. Por ejemplo, las glaciaciones más recientes (ocurridas con periodicidades diversas, básicamente entre 20 mil y 400 mil años) tuvieron también una importancia y una alta correlación en fase, con la variación del $CO_2$ atmosférico (9). Las causas de esta correlación son pobremente entendidas.

Sin entrar en otros detalles paleoclimáticos, este escenario nos muestra una atmósfera sujeta a perturbaciones externas, con variaciones cíclicas y esporádicas, lejos de la visión actual de un mundo equilibrado, aunque frágil y propenso a cambios irreversibles producto de la acción humana. Pero además, muestra su capacidad de adaptación al cambio, de complejidad creciente, de emergencia de propiedades sorprendentes, generadas ante el desequilibrio y las perturbaciones fortuitas (inclusive catastróficas); todas estas cualidades, inherentes a la sostenibilidad como materia de estudio (2), se ven claramente expresadas en el mundo natural, y no solo en el biológico, también en el *sistema climático*.

El clima terrestre en absoluto ha permanecido constante a lo largo de las eras geológicas, ni siquiera en tiempos históricos. Las glaciaciones son una muestra de esta variabilidad natural del sistema climático*,* en este caso, totalmente controladas por la orientación del eje de rotación terrestre respecto a un sistema inercial, fijo en el espacio. Más recientemente, se ha reconocido durante el último milenio la existencia de un *Óptimo Medieval* (alrededor del siglo XII) con temperaturas muy cercanas a las medias actuales (10); en esta época los Vikingos establecieron colonias en América; posteriormente, las reconstrucciones climáticas y datos dendrocronológicos (11) dan cuenta de una *Pequeña Edad de Hielo* (LIA, por Little Ice Age, en inglés): por ejemplo, el río Támesis solía congelarse (episodio ocurrido por última vez en 1895). La LIA fue acompañada por grandes mínimos en la actividad solar. Los Grandes Mínimos, son mínimos prolongados de diferente duración en la actividad magnética del Sol, evidenciados como ausencia de manchas solares ((12) y referencias allí incluidas). Los más estudiados de esos eventos, el Mínimo de Maunder (MM, 1645-1712) y el Mínimo de Dalton (DM, 1790- 1830), fueron períodos de gran disminución en el número de manchas solares, acompañados de apreciables caídas en la temperatura planetaria media, lo cual ha sido confirmado por datos arqueológicos y recientemente por reconstrucciones a partir de "proxy-datos" indicadores de anomalías en la temperatura media global (11). También fueron épocas de gran actividad volcánica: en 1815 la erupción del Pinatubo, emitió enormes cantidades de aerosoles que produjeron un forzante radiativo de signo negativo en el clima, lo que originó un descenso anual de la temperatura en todo el planeta, 1815 es comúnmente conocido como "el año sin verano". En Sudamérica, la LIA se presume un período frío y seco (13); coincidente con los hallazgos de miles de cadáveres de vacunos pertenecientes a ésa época, encontrados en la zona Pampeana de Argentina (y que Charles Darwin asentara en las notas de su celebrado viaje).

Varios autores han propuesto que la actividad solar es, en realidad, la responsable de estas variaciones o mínimos climáticos (14). Las posibles relaciones entre el clima, las manchas solares y el vulcanismo se estudian desde hace mucho tiempo (7). Sobre todo, las manchas solares presentan una correlación bastante clara con la temperatura media, fenómeno que todavía no se comprende muy bien. Los glaciares por ejemplo, cuya retracción masiva comenzó *antes de la industrialización*, no muestran signo de dependencia muy alta con el calentamiento global (8, 15). En Argentina, la mayor parte del país ha soportado en las últimas décadas un cambio en el régimen hídrico a gran escala, presentándose más lluvioso a partir de la década de 1950-60 ((16), (17) y referencias allí incluidas), esto ha provocado el corrimiento de la frontera agrícola hacia el Norte – Noreste (17), lo cual ha sido altamente satisfactorio para el sector agrícola del país. Aquí es inevitable mencionar el fenómeno de El Niño-Oscilación Sur (ENOS), la modificación acoplada de las corrientes oceánicas y los vientos alisios del Pacífico Sur. Especialmente después del evento de los años 1982-1983 cobró gran interés en los sectores de la economía y la tecnología (18) debido a sus devastadoras consecuencias globales. Esto eventos, que son fuente "inmediata" de variabilidad climática (principalmente extremos en las condiciones del clima) manifiestan que los procesos atmosféricos con respuesta de corto período son, en determinado contexto, más importantes que los escenarios a futuro, ya que tienen consecuencias enormes y concretas para la sociedad y la economía mundial ((2), (18), y referencias allí incluidas). Por lo tanto, lo expresado en los párrafos precedentes, nos indica que el clima posee una propiedad intrínseca de *variabilidad natural*.

Aun si los parámetros atmosféricos promedio no tuvieran variaciones estadísticamente significativas en un determinado sitio, el problema del transporte y la deposición de contaminantes atmosféricos requieren de un conocimiento adecuado de, al menos, los vientos dominantes, su interacción con las irregularidades del terreno y la estabilidad atmosférica (movimientos verticales de la masa de aire) en un determinado día. Si se desea un conocimiento más allá del modelo de "caja negra" o entender procesos de contaminación a escala sinóptica (miles de kilómetros), la comprensión básica de los grandes movimientos horizontales de la masa de aire es imprescindible. Todas las actividades generan cambios en el entorno y estos se propagan en mayor o menor medida mediante teleconexión y dinámica caótica; así, resulta claro que el conocimiento de la meteorología relevante en los procesos de contaminación del aire puede abordarse en distintas escalas de complejidad y su estudio resulta indispensable para los interesados en el control y mitigación de la contaminación atmosférica.

Extenderse en todos los temas inherentes a estas problemáticas está muy por fuera del alcance de este trabajo; podemos terminar esta revisión teórica planteando que el profesional de la ingeniería, empático con el medioambiente y con sus conciudadanos, debería estar en condiciones de estimar las consecuencias de su trabajo; debería al menos, en principio, tener una formación básica que le permita ensayar respuestas a preguntas como: ¿qué tipo de emisiones estoy produciendo?, ¿cuál es la distribución de tamaños de partículas emitidas en el proceso?, ¿qué probabilidad de recibir partículas menores que 10 µm tienen vecinos a 10 cuadras; y a 10 km?, ¿de qué parámetro meteorológico depende el arrastre de determinado contaminante gaseoso?, ¿cómo se comporta ese parámetro en la zona? ¿Cuáles podrán ser las consecuencias en una obra a la intemperie que involucra grandes remociones de tierra, si es año de La Niña, El Niño o de fase neutral?, ¿qué relación podrá existir con el mes del año? Habiendo trabajadores que realizan su tarea al aire libre, ¿existe un riesgo cancerígeno mayor actualmente, debido a la exposición solar, que hace 50 años? ¿Las emanaciones gaseosas con las que trabajo, pueden llegar a cambiar el clima?, ¿existen otras causas más probables de cambio climático en mi entorno, como por ejemplo el metano emitido por rumiantes?, ¿conviene que llueva, la lluvia "limpia la atmósfera"?, ¿tendremos lluvia ácida por aquí? Si estoy emitiendo *x* cantidades de $CO_2$ a la atmósfera, ¿qué relación podrá tener con una hectárea de maíz, sembrada a 500 m o a mayor distancia? ¿Cuál es el efecto de la radiación solar sobre los contaminantes o respecto a la formación de ozono troposférico?, siendo la radiación solar la principal fuente de energía para la atmósfera o para la edafología, ¿qué hay del recurso solar en la zona?, ¿cómo puede cuantificarse y modelarse? Durante años de gran actividad solar, ¿qué tipo de problemas pueden esperarse en determinado dispositivo de telecomunicaciones terrestres y satelitales? ¿Y en subestaciones de potencia, qué tipo de problemas pueden aparecer en la distribución de energía eléctrica? ¿Y que hay respecto a la energía solar como fuente renovable más evidente?

Una materia basada en Física de la Atmósfera debería proporcionar aptitudes profesionales para dar respuestas a preguntas más o menos básicas como las formuladas y elementos para tratar con cuestiones más avanzadas. Debería comenzar por explica la meteorología observada a escala local y zonal; que es la relevante a los problemas de dispersión de contaminantes. Una vez que se tienen estas herramientas, estamos en condiciones de evaluar el transporte y la deposición de contaminantes sólidos y gaseosos en la atmósfera (y de utilizar los modelos y programas computacionales que son *estado del arte* en el tema).

El Sol, siempre presente en estos temas, es la principal fuente de energía y de él dependen muchos fenómenos meteorológicos y climáticos (por ejemplo, la formación de ozono como contaminante secundario). La atmósfera produce absorción (destrucción de fotones) y scattering (dispersión o transformación de fotones, a otras longitudes de onda) y variación de la heliofanía (horas de Sol), de las cuales a su vez depende el recurso solar en tierra. Además, la interacción de la atmósfera con la radiación solar incidente, mediante la magnetósfera, tiene consecuencias para los ingenios humanos; por ejemplo, disturbios en la distribución de energía eléctrica (19). Dicho esto, es inmediato pensar en introducir temas relacionados con la energía solar, como la cuantificación y evaluación del recurso solar en determinada zona y su relación con sus *usos concretos* y como *fuente de energía renovable*.

Por último, podemos adentrarnos en cuestiones más generales del clima, como problema físico, medioambiental y también social. Respecto a este último punto, durante la última década el cambio climático ha comenzado a ser puesto en un contexto económico. El *informe Stern* ha generado alarma en el mundo de la economía (2); sin embargo, los indicadores económicos clásicos (como el producto bruto interno) aplicados a escenarios a futuro con pronósticos altamente negativos, pueden resultar totalmente ajenos a las verdaderas necesidades de la población (20). Esto nos muestra que el cambio climático ha trascendido las fronteras de las ciencias naturales, para adentrarse en otros campos de gran incumbencia para los ingenieros.

## 3. OBJETIVOS DE LA MATERIA PROPUESTA

En base a los ejes temáticos desarrollados y a las cuestiones de índole teórico-práctico anteriormente planteadas, los objetivos de la materia propuesta ciertamente pueden presentarse como generales y específicos. Un objetivo general es desarrollar aptitudes para tender a la comprensión del mundo como *un todo*, lo cual es fundamental para la consideración simultánea dentro de la práctica profesional del ingeniero, de los sistemas sociales, económicos y medioambientales, y la viabilidad a largo plazo entre estos sistemas.

Más específicamente, intentamos completar la formación científico-tecnológica en este campo del conocimiento para comprender mejor: procesos a micro y meso-escala atmosférica; la estratificación térmica de la atmósfera y su impacto en la dispersión de contaminantes gaseosos y sólidos, y en el establecimiento del clima terrestre; el efecto del Sol en la meteorología, el clima y como recurso energético concreto y mensurable; la comprensión de lo que es cambio y variabilidad climática; etc. Otro punto específico es el manejo de la información disponible, de los datos, especialmente meteorológicos, con sus alcances y limitaciones. Por otra parte la identificación de sistemas locales o zonales de importancia medioambiental o actividades de gran impacto ambiental, de tipo natural o agro-industrial.

## 4. CONTENIDOS MÍNIMOS PROPUESTOS

Proponemos el siguiente conjunto de contenidos mínimos, los cuales, así distribuidos, cumplen además con el objetivo de introducir gradualmente los temas partiendo de aquellos con mayor base observacional y práctica; por eso, el tema Energía Solar (que a pesar de ser tratado en función de sus efectos sobre la atmósfera, el medioambiente y el clima, lo presentamos como capítulo aparte), es mejor desarrollarlo *antes* que dispersión en la atmósfera. De esta manera, la materia se va concatenando de tal suerte que resulta gradual la conceptualización de temas cada vez más abstractos o con mayor rigor matemático.

**Capítulo 1 – Introducción a la atmósfera terrestre**
> Introducción. Características globales: masa, distribución. Composición.
> Origen y evolución primigenia. Fraccionamiento y gases nobles. Efecto de los eventos catastróficos.
> Estructura general y estratificación térmica. Diatermancia. Cambio de la densidad con la altura y la temperatura. Movimientos verticales. Gradientes térmicos. Estabilidad. Inestabilidad. Neutralidad. Capa inversora.
> Movimientos horizontales: Circulación general. Celdas. Alisios. Consecuencias. Vientos.
> Condensación atmosférica. Lluvia, nieblas y tormentas.
> Pronósticos e información meteorológica.

**Capítulo 2 – Energía solar (irradiancia y magnetismo)**
> El Sol: estructura y funcionamiento. Magnetismo, dínamo y tachoclina.
> Irradiancia solar. La "constante" solar y su variación. Energía solar en superficie.
> Fórmulas prácticas-simples para el cálculo de la radiación solar, horario de salida y puesta y cantidades derivadas.
> Radiación UV. Ozono troposférico y estratosférico. Agujero de ozono austral y boreal.
> Actividad solar y ciclos del Sol. Medioambiente y clima espacial.
> Magnetosfera terrestre, efecto solar en las perturbaciones geomagnéticas.
> Tormentas solares. Posibles efectos sobre el medioambiente, los artefactos y las telecomunicaciones.
> Probabilidad de ocurrencia de eventos. Parámetros indicadores. Sistemas de información.

**Capítulo 3 – Dispersión y transporte en la atmósfera**
- > Altura de mezcla. Descripción de la dispersión de una columna de humo. Pluma Gaussiana.
- > Dispersión de contaminantes: difusión y transporte. Longitud de Monin-Obukhov. Parametrización horizontal del viento. Velocidad y escala de fricción. Métodos eulerianos y lagrangianos. Ecuaciones básicas y sus soluciones.
- > Modelos de dispersión de contaminantes industriales. Simulaciones. Programas de cálculo de la EPA (Environmental Protection Agency).
- > Material particulado. $PM_{10}$. Material particulado grueso. Modelos y Simulaciones.
- > Problemática de la actividad agropecuaria. Cenizas volcánicas. Naturaleza, dispersión e impactos en la sociedad.
- > El Sol y los contaminantes secundarios. Lluvia ácida. Isla Térmica.

**Capítulo 4 – Clima**
- > El clima de un lugar. Forzantes internos y externos al planeta.
- > El clima en la geología y la historia.
- > El fenómeno ENSO (El Niño- Oscilación Sur).
- > Modificación natural y artificial del clima. Efecto invernadero y el ciclo del $CO_2$. La teoría del cambio artificial del clima.
- > El río Paraná y su relación con los forzantes del clima. Crecidas y bajantes. Relación con ENSO, manchas solares y movimiento solar baricéntrico.
- > Variabilidad climática. Glaciaciones. Teoría de Milankovitch.
- > Pequeña edad de hielo. Calentamiento medieval. Grandes Mínimos. Teoría del aumento del $CO_2$.
- > Calentamiento global o cambio climático. Proyecciones y controversias.
- > Nociones de agroclimatología. Sequías. Heladas. Crecimiento de los cultivos con relación al $CO_2$.

De los temas propuestos, queda claro que la articulación será básicamente vertical y se dará con materias de diversos años y especialidades, del tipo básico, como Física General (I y II) y Análisis Matemático I y II, y otras más específicas, como Higiene y Seguridad, Ingeniería Ambiental, Estadística, Análisis Numérico y Cálculo Avanzado. En particular será campo propicio de aplicación de métodos numéricos estudiados en estas últimas materias, en años superiores, y de profundización en temas como análisis de Series de Tiempo, Modelado y Simulación de Procesos.

Respecto a la aprobación de la cursada, resulta interesante plantear trabajos finales totalizadores, que sean de aplicación concreta, dependiendo de la especialidad (tipo de ingeniería) y de la zona donde se dicte la materia (enfoques *regionales* para el análisis de forzantes del clima y variabilidad climática). En nuestro caso, la materia propuesta está concebida para ser dictada en una Regional de UTN situada en la zona norte de la provincia de Buenos Aires. En esta zona, es imprescindible presentar los impactos climáticos en la dinámica del río Paraná, tener buena práctica sobre dispersión de contaminantes en complejos industriales y dar elementos respecto a la problemática ambiental agrícola; entre otras cosas, para evaluar sus posibles efectos sobre el medioambiente circundante. Estos temas son interesantes para realizar trabajos grupales, donde se ejercite además, la metodología de la investigación científica: búsqueda bibliográfica relevante, estado del arte, temas abiertos, preguntas relevantes, generación de hipótesis, etc. Por ejemplo, la estimación de posibles inundaciones relacionadas al río Paraná, en un período determinado de tiempo en el cual se ha proyectado determinada obra, puede ser un buen problema de aplicación; para esto es necesario recabar la información en los centros especializados (series de alturas del río, índices específicos de ENSO, y de los forzantes externos (astronómicos) involucrados, etc.). Un trabajo de prognosis en este campo puede resultar muy útil, estimulante y fácilmente realizable. Otro trabajo de aplicación, más orientado a eléctricos y electrónicos, puede ser determinar las *corrientes geomagnéticas inducidas* y su relación con la distribución eléctrica, debidas a un pronóstico de gran actividad solar. Por supuesto, existen aplicaciones concretas a problemas más comunes como determinar la estructura de una pluma gaussiana emitida en determinado proceso industrial, o la zona de caída de material particulado liberado en una acería (de mayor interés para ingenieros industriales y metalúrgicos). Los trabajos práctico también pueden incluir resolución de problemas (aunque muy concretos y dirigidos); por ejemplo, en al Cap. 1 pueden proponerse problemas de aplicación del Teorema de Bernoulli para determinar cuándo un planeta retiene o libera atmósfera, como así también cálculos referidos a la meteorología local de un sitio; en el Cap. 2 pueden calcularse diversos parámetros de radiación solar (irradiancia diaria a partir de valores de insolación; análisis de correlación lineal entre la radiación solar diaria y la amplitud térmica local); en el Cap. 3 pueden estimarse determinados parámetros aerodinámicos referidos a la turbulencia y las características del terreno, además de problemas de difusión los cuales son en realidad más ricos de plantear computacionalmente. Por último, los temas del Cap.4 permiten realizar análisis estadísticos y de series de tiempo a diversas variables climatológicas y físicas relacionadas.

## 5. COMENTARIOS FINALES Y CONCLUSIONES

Como materia, Física de la Atmósfera puede remontarse a obras seminales como *Physic of the Air* de H. Humphrey ((8), con una primera edición de 1924), quien también profundizó en las cuestiones de variabilidad climática, relación con el Sol, vulcanismo, etc. Aunque, en general, no existe mucha bibliografía en las bibliotecas de las facultades de ingeniería respecto a estos temas, en internet existe profusa información, artículos (especialmente los suministrados por la biblioteca virtual del Ministerio de Ciencia y Tecnología de Argentina) y publicaciones de instituciones relevantes a nivel mundial (CEPAL, etc.), sólo es cuestión de buscar un poco. Otro libro fundamental, minuciosos y extenso, relacionando física atmosférica con problemas concretos de contaminación es *Atmospheric Chemistry and Physics of Air Pollution*, de J. Sheinfeld (21).

La propuesta que aquí presentamos es original y el curso no ha sido implementado todavía. Para una presentación más concreta de los trabajos prácticos es inevitable esperara la experiencia de una cursada, sobre todo por tratarse de una materia para todas las especialidades los trabajos prácticos deberían adaptarse a las especialidades de los alumnos. En base a todo lo expuesto y al programa tentativo presentado, concluimos que es posible, aun en un cuatrimestre, formar ingenieros que comprendan los procesos básicos que dieron origen a la atmósfera actual (muchos de ellos catastróficos), los procesos atmosféricos locales y globales relacionados con el clima, la dispersión de contaminantes atmosféricos, y conceptos clave como actividad solar, cambio y variabilidad climática. El cambio climático puede exponerse de forma crítica, analizándose las fuentes más antiguas y mejor conocidas de variabilidad climática como ENOS y el impacto que tienen en la actualidad y han tenido en épocas pasadas.

La clave evolucionaria del mundo natural es el desequilibrio y la inestabilidad; sin ellas los organismos no evolucionan, y la Tierra no hubiera sido lo que hoy es, un planeta único en el Universo conocido. La visión *goreana* (por Al Gore, con relación a su famosa película "*Una verdad incómoda*") de caos inminente puede ser útil para una toma de conciencia de los problemas ambientales y sus posibles consecuencias; pero, sin dudas, no contribuye al real conocimiento de la naturaleza y su interacción con el mundo artificial, pilares en el paradigma de la sostenibilidad. Poner en contexto el calentamiento global actual en la historia y el paleoclima y la variabilidad climática natural en el mundo sostenible es fundamental para la comprensión del mundo natural. Aun si las consecuencias del cambio climático fueran positivas, contaminar y emitir gases y sólidos sin control es altamente perjudicial para todos. La atmósfera, desde una óptica económica, es un bien público y, en este sentido, contaminarla o alterarla significativamente en su funcionamiento representa la mayor externalidad negativa posible (2).

De esta forma, pueden presentarse los procesos de contaminación y el cuidado del medioambiente más allá de un contexto meramente descriptivo basado en la hipótesis del daño "antropogénico"; sino más bien, fundamentados en la física de los procesos naturales y las posibles rutas dinámicas propias del sistema climático.

## Agradecimientos

## Referencias